# Tensorial origin of the action function, nexus between Quantum physics and General Relativity

Enrique Ordaz Romay[1].

*Facultad de Ciencias Físicas, Universidad Complutense de Madrid*

## Abstract

The analytic physics, when it is development from aprioristic form, constructs all the laws from the Hamilton principle, also called action principle. According to this principle all systems are characterized by a magnitude called action which variation is equal to zero for the real evolution of the systems. At first, do not exist restrictions for the mathematical form of the action function, nonetheless, as in the a posteriori developer of the analytic physics, the traditional equations of the physics lead up to a scalar action, it is always sure that the action may be a scalar function. In this article we propose the existence of a tensorial function like origin of the traditional scalar action. First because it permits, without other argumens, to justify the own action principle and second because it permits connect the complementary principle of the Quantum physics with the analytic expressions of the nonquantum physics, including the General Relativity. We call tensorial action to this original function and their components are connecting, directly, with the tensorial (spinorial) componets of the wave function of the particles with spin higher than zero. For this reason we can understand why in the nonquantum physics only has significance a scalar action.

---

[1] eorgazro@cofis.es

# Tensorial origin of the action function.

All physical theory, either classic, relativistic or quantum, mechanics of material points, of solid or of fluids, of fields, etc., can be developed begining from a single principle well-known like the Hamilton´s principle or action principle. Its enunciated for the mechanics is [1, §2]: All mechanical systems are characterized by a function $L = L(q,\dot{q},t)$ called lagrangian function and the real movement of the system satisfies the following condition:

$$\delta S = \delta \int_{t_1}^{t_2} L(q,\dot{q},t)dt = 0 \qquad (1)$$

This enunciated is completely similar for any physics branch. In more general way we can rewrite this enunciated in the next form: "All physical systems are characterized by a magnitude *S* called action which is defined by $S(t) = \int L(q,\dot{q},t)dt$ and for a real evolution of the system $q = q(t)$ takes a extreme value" [2, §8].

The different physical theories can be built starting from this principle only give a concrete mathematical expression for the lagrangian and the generalized coordinateses. Likewise, if one makes the variation (1) we obtained Euler-Lagrange equations that to the being resolved show the physical properties of the system.

All physical property of any system, just as it can be: the second Newton law, the energy conservation principle, the light trajectory, the Schrödinger equation or the Dirac one, the Bernouilli fluids equation, etc, all of they can be deduced by a action principle when the action adopts the appropriate expression, in according to the mathematical apparatus that is used in such a theory.

In this form, the action principle is the last and the only explanation for all physical property of a system. The whole physical science decreases to a single principle although this has several mathematical realizations depending of the branch of the physics that is applied.

But, why should all physical systems obey the action principle?, that is to say, why should all physical systems be characterized by a magnitude which becomes extreme? The explanation is not in the physics, but in the math[2].

Let us suppose that we have a tensorial function $S^{ijk...}$ which represents mathematically to our system.

According to a well-known theorem of the tensorial calculation [3 §3.3]: the scalar product $A^i B_i$ is invariable before a transformations of coordinated.

This result can be generalize easily: Let $\{x^i\}$ and $\{x'^i\}$ be two systems of coordinated. The transformation from a system of coordinate to the other one, for a tensile covariant, has the form: $A^{ijk...} = \frac{\partial x^i}{\partial x'^l}\frac{\partial x^j}{\partial x'^m}\frac{\partial x^k}{\partial x'^n}\cdots A'^{lmn...}$. The scalar product of two tensors with same number of index will be then:

$$A^{ijk...}B_{ijk...} = \frac{\partial x^i}{\partial x'^l}\frac{\partial x^j}{\partial x'^m}\frac{\partial x^k}{\partial x'^n}\cdots\frac{\partial x'^r}{\partial x^i}\frac{\partial x'^s}{\partial x^j}\frac{\partial x'^t}{\partial x^k}\cdots A'^{lmn...} B'_{rst...} =$$

$$= \frac{\partial x'^r}{\partial x'^l}\frac{\partial x'^s}{\partial x'^m}\frac{\partial x'^t}{\partial x'^n}\cdots A'^{lmn...} B'_{rst...} = A'^{lmn...} B'_{lmn...}$$

Applying this result to the scalar product of one tensor times itself:

$$S^2 = S^{ijk...} S_{ijk...} = \text{constant} \qquad (2)$$

When makes a variation of a magnitude, it is varying in relation to the objects which has same variables in this magnitude. In the case of tensors whose indexes reference to the coordinates on which makes the variation, this variation go with the corresponding transformation of the coordinates. According to this, by virtue of (2) be deduce:

$$\delta(S^2) = 0 \qquad (3)$$

---

[2] The Feynman method of the path integral, although it allows to deduce the Hamilton principle of the classic physics or the Schödinger equation of the quantum physics, it is not in itself a demonstration, because it comes to substitute an axiom (the Hamilton principle) for another axiom (the call Feynman's principle of the democratic equality of all histories).

The scalar product of two tensors is, indeed, a scalar. Therefore, in the scalar product of a tensor times itself the result it is a square scalar, this is $\underline{S}^2$ (starting from here it will be denoted the scalars with underlined).

$$S^2 = \underline{S}^2 = \text{scalar constant}$$

Substituting this result in (3) it is obtained $\delta(\underline{S}^2) = 0$ that to being the square of a scalar can be operate until to obtain

$$\underline{S} \cdot \delta(\underline{S}) = 0$$

For check this relationship it should be made zero at least one of the two multiplicand, that is to say or $S = 0$ or $\delta \underline{S} = 0$. But, if the first one is completed the second one is always completed, then in anyway to be completed that:

$$\boxed{\delta \underline{S} = 0} \tag{4}$$

In consequence: if the tensor $S^{ijk...}$ characterizes to a physical system completely, then it is verify the relationship (4) which is identical to the Hamilton principle (1).

If we consider the existence of a tensorial function which characterizes to the system completely, all the physical properties of the system are deduced mathematically from the formalism which is used to represent the physical objects.

## The tesorial action in quantum.

In quantum, the correspondence principle [78 §7] takes to the expresion:

$$\hat{E}(\Psi) = -i\hbar \frac{\partial \Psi}{\partial t} \quad ; \quad \hat{p}_x(\Psi) = i\hbar \frac{\partial \Psi}{\partial x} \quad ; \quad \hat{p}_y(\Psi) = i\hbar \frac{\partial \Psi}{\partial y} \quad ; \quad \hat{p}_z(\Psi) = i\hbar \frac{\partial \Psi}{\partial z}$$

This expression is valid so much in nonrelativistic quantum [5 §15] like in the relativistic [6 §2.1]. It can express in 4-vectorial notation in a simple form making the same process used in special relativity:

$$P_0 = \frac{E}{c} \quad ; \quad P_1 = -p_x; \quad P_2 = -p_y; \quad P_3 = -p_z \quad \Rightarrow \quad \hat{P}_i(\Psi) = -i\hbar\frac{\partial \Psi}{\partial x^i} = -i\hbar\partial_i\Psi \tag{5}$$

On the other hand, when the action is a physical magnitude dependent of the coordinateses, adopts, so much for the mechanics of material points [1 §43], like for the fields theory [2 §9], the form:

$$P_i = \frac{\partial S}{\partial x^i} = \partial_i S \tag{6}$$

Comparing (5) and (6) is obtain $\partial_i S = -i\hbar\partial_i \Psi$ which operating plays to:

$$S = -i\hbar\Psi \tag{7}$$

The development utilized for arrive to the expression (7) is relativistic invariant just as it guarantees all the equations utilized.

In relativistic quantum the wave functions are tensorial according to the spin of the system. That is to say, for spin 0 the wave function is a scalar, for spin ½ it is vectorial (spinorial) [6 §3.3], for spin 1 it is a matrix [6 §5.1] and in general case, for spin n/2 the wave function will be a n indexes tensor [6 §5.3]. Applying this concept to the expression (7) is obtained:

$$\boxed{\hat{S}_{ijk...}(\Psi_{ijk...}) = -i\hbar\Psi_{ijk...}} \tag{8}$$

That is to say, it is necessary to adopt that the operator action is tensorial to guarantee the tensorial character of the expression.

The wave function, just as it is represented in the expression (8), is not identified exactly with the Dirac wave function. For the Dirac wave function the scalar $\Psi^2$ has the form $\Psi^+\gamma_0\Psi$ being $\gamma_0$ the gamma zero matrix. However, the expression (8) plays to a to scalar with the form $\Psi^2 = \Psi^+ G\Psi$ being $G$ the metric matrix, whose properties are very different to properties of the gamma zero matrix.

To transform some wave functions in the other ones first we should restrict the expression (8) to the case of the Minkowski space and later to compare the scalar products. Making this it is obtained, for the case of spin ½, that all the components are equal except the second one which will have the form $\Psi^1_{Dirac} = -\Psi^1_{vectorial}$.

The expresion (8), as well as can to demonstrate the action principle in quantum (because it is a tensorial action) also it allows to deduce the most important principles of the quantum physics, besides to explain the reason why the physics for the present has not applied the magnitude action like a tensor.

**The overlapping principle.**

Let us suppose two independent and isolated systems 1 and 2, characterized by their corresponding tensorial actions $S_1$ and $S_2$ (starting from here the notation used will be without indexes for the tensors, for further clarity, all the magnitudes in capitals will be tensorials, the scalars will be denoted by minuscule or underlined capitals. In this form, the superindexes will denote exponents and the subindexes will refer the system to which refer the expression). The sum of both actions, multiplied for constant $a$ and $b$ has the form $S = aS_1 + bS_2$. Its square is:

$$S^2 = a^2 S_1^2 + b^2 S_2^2 + 2\operatorname{Re}(ab) S_1 \cdot S_2 \qquad (9)$$

When the systems 1 and 2 is orthogonal, that is to say, they are different stationary states of the same system, the product to climb $S_1 \cdot S_2$ is made zero and the expression (9) reduce to $S^2 = a^2 S_1^2 + b^2 S_2^2$.

As $\delta(S_1^2) = 0$ and $\delta(S_2^2) = 0$ then from the last expression is deduced automatically that $\delta(S^2) = 0$. That is to say, the tensor $S = aS_1 + bS_2$ completes the action principle. Applying (8) it is obtained:

$$\boxed{\Psi = a\Psi_1 + b\Psi_2} \qquad (10)$$

which is known as the states overlapping principle [5 §2].

**The Born´s probabilistic principle.**

If one makes the variation of the general expression (9), to bear in mind $\delta(S^2) = 0$, $\delta(S_1^2) = 0$ and $\delta(S_2^2) = 0$, it is obtained $\delta(S_1 \cdot S_2) = 0$.

The last expression coincides with the action principle again, for a form of the action similar to $S_{12} = S_1 \cdot S_2$. Substituting here the expression (8) it is obtained:

$$\Psi_{12} = \Psi_1 \cdot \Psi_2 \qquad (11)$$

This last expression coincides with the action principle again, for a form of the action similar to $S_1 \cdot S_2$. As this product of tensorial actions is like a scalar, this indicates us this product should coincide with a square tensorial action, that is to say: $S_1 \cdot S_2 = S_{12}^2$, Substituting here the relationship (8), it is obtained:

$$\Psi_{12}^2 = \Psi_1 \cdot \Psi_2 \qquad (11)$$

This expression play to the Born´s probabilistic principle [4 §11] since it interprets the wave function which characterizes to a composed system as the product of the singular wave functions, the same way as a probabilistic magnitude work [5 §2].

**The condition of normalization of the wave function.**

The expression (8) it drives directly to $S^2 = -\hbar^2 \langle \Psi | \Psi \rangle = \text{constant}$ (here we use the Dirac´s notation), or the same thing:

$$\int_{-\infty}^{+\infty} \Psi(q)\Psi^+(q)dq = \text{constante} \qquad (12)$$

This expression indicate the integral should be convergent and therefore the function $\Psi(q)\Psi^+(q)$ behaves like a probability distribution. As the lineality of the wave function, the constant is arbitrary and we can fix it for usefulness. When we make this constant similar to 1, the expression (12) takes the well-known form like normalization condition [5 §2].

## The action like scalar magnitude in the traditional physics.

The expression (8) also explains why the tensorial action is not used in classical physics, relativistical physic, nonrelativistical quantum and even in relativistical quantum.

Firstly, in classical physics and in relativistical physic the concept of spin doesn't exist, or in another form, always the spin is zero in no-quantum systems [5 §54]. Therefore in these systems the action is scalar.

On the other hand, the norelativistical quantum is developed basically for particles without spin [5 §54]. No will be up to 1925 that G. Uhlenberk proposes the existence of an intrinsic angular momentum for the electron like explanation of the call effect anomalous Zeemann [7 §§3.6 and 3.7] and in 1927 W. Pauli introduces the spin in the formalism of wave functions [5 §54] as a special vector, calls espinor, which multiplies to the traditionally scalar wave function [5 §55]. Due to this method to introduce the spin, the nonrelativistical quantum continue deducing from a scalar action.

The case of the relativistical quantum mechanics is different. Although the Dirac´s equation, for example, is obtained when solving the relativistical expression of the energy [6 §3.1], the result that one obtains it is vectorial (espinorial) [6 §3.3]. However, the usage, a priori, of the Minkowski metric [6 §1.4] as geometry of the 4-space in relativistical quantum makes unnecessary the tensorial character of the action (in this metric the space is plane and $R^i_{klm} = 0$) to exception of we want to demonstrate the action principle or the substitution principle for potential in the wave equation [6 §7.8] both things outside of the current physics just as we knew it.

# Acknowledgement:

In the first instance, I thank Juan Ramón Péris for his backing, help and advice. Thanks to Jesús Angulo for takes an interest. Thanks to Juan Ramón Knaster from CERN for his patience and for liven up me for publicate. Finally, thanks (again) to Juan Ramón Péris and José Fernando Señarís for revise the last version of this paper.

# References

[1] L. D. Landau & E. M. Lifshitz, *Mechanics*, 2ª ed. (Addison-Wesley. 1965).

[2] L. D. Landau & E. M. Lifshitz, *The Classical Theory of fields*, 2ª ed. (Addison-Wesley, 1967).

[3] D. C. Kay, *Tensor Calculus* (MacGraw-Hill, 1988)

[4] A. Galindo & P. Pascual, *Mecánica cuántica* (Alhambra 1978).

[5] L. D. Landau & E. M. Lifshitz, *Quantum Mechanics (nonrelativistical theory)*, (Addison-Wesley 1963).

[6] F. J. Yndurain, *Mecánica cuántica relativista*, (Alianza 1990).

[7] M. Alonso & E. J. Finn, *Quantum and Statistical Physics* (Addison-Wesley, 1968)